\documentclass[pre,nofootinbib,superscriptaddress,showpacs,showkeys]{revtex4}

\usepackage{graphicx,epsfig}
\usepackage{amsfonts,amsmath,amssymb,amsthm,amscd}
\usepackage[T2A]{fontenc}
\usepackage[cp1251]{inputenc}
\begin{document}

\title{Directed Coulomb Explosion regime of ion acceleration from mass limited targets
by linearly and circularly polarized laser pulses.}

\author{S. S. Bulanov}
\affiliation{FOCUS Center and Center for Ultrafast Optical Science,
University of Michigan, Ann Arbor, Michigan 48109, USA}
\affiliation{Institute of Theoretical and Experimental Physics,
Moscow 117218, Russia}

\author{D. W. Litzenberg}
\affiliation{Department of Radiation Oncology, University of
Michigan, Ann Arbor, Michigan 48109, USA}

\author{K. Krushelnick}
\affiliation{FOCUS Center and Center for Ultrafast Optical Science,
University of Michigan, Ann Arbor, Michigan 48109, USA}

\author{A. Maksimchuk}
\affiliation{FOCUS Center and Center for Ultrafast Optical Science,
University of Michigan, Ann Arbor, Michigan 48109, USA}

\begin{abstract}
We consider the Directed Coulomb Explosion regime of ion
acceleration from ultra-thin double layer (high Z/low Z) mass
limited targets. In this regime the laser pulse evacuated the
electrons from the irradiated spot and accelerates the remaining ion
core in the direction of the laser pulse propagation. Then the
moving ion core explodes due to the excess of positive charge
forming a moving charge separation field that accelerates protons
from the second layer. The utilization of the mass limited targets
increases the effectiveness of the acceleration. It is also shown
that different parameters of laser-target configuration should be
chosen to ensure optimal acceleration by either linearly polarized
or circularly polarized pulses.
\end{abstract}

\maketitle

\section{Introduction}

{\noindent}The acceleration of ion beams in the intense laser pulse
interaction with gaseous and solid density targets is one of the
most promising applications of future laser systems \cite{multiMeV}.
These beams can be used in hadron therapy \cite{hadron therapy},
fusion ignition \cite{FI}, proton radiography \cite{Borghesi}, and
for injection into conventional accelerators \cite{accelerator}. The
possibility of effective ion acceleration was proved in a number of
experiments, which showed the maximum ion energy up to tens of MeV.
This process also was extensively studied analytically and in
particle-in-cell (PIC) computer simulations, which showed the
possibility of ion acceleration up to the energy of several hundred
of MeV or even to the several GeV depending on the parameters of
laser-plasma interaction \cite{2d3d,2d3d-1}.

There are a number of ion acceleration regimes discussed in the
literature: (i) Target Normal Sheath Acceleration (TNSA)
\cite{TNSA}; (ii) Coulomb Explosion (CE) \cite{CoulombExplosion};
(iii) Radiation Pressure (RPDA) \cite{LP}; and (iv) acceleration
from near critical density targets (NCD) \cite{NCD}. The TNSA regime
is the first identified with regard to ion acceleration from thin
foils. It comes into play when the laser pulse is not intense enough
to penetrate the target and instead it launches hot electrons from
the front of the target that go through the target and establish a
sheath field at its back. This field accelerates the ions. When the
pulse is intense enough to evacuate all the electrons from the focal
spot the Coulomb Explosion of the bare ion core follows. And when
the pulse is able to push the foil as a whole this is the Radiation
pressure regime of ion acceleration. The TNSA, CE and RPDA regimes
use thin foils of solid density as targets, in contrast to that the
NCD regime utilizes targets of near critical density. The ion
acceleration in this case is due to the longitudinal electric field
established at the back of the target by the exiting the propagation
channel magnetic field, which in its turn is generated by electrons
accelerated in the wake of the laser pulse. Recently several
mechanisms of ion acceleration were reported in the literature: the
TNSA coupled to the energy conversion from an expanding electron
cloud to the expanding ion cloud under the action of burning through
the foil laser pulse \cite{Hegelich}, the CE optimized by injecting
a proton bunch into the longitudinal field \cite{Ma2007}, and the
Directed Coulomb Explosion (DCE), which is the effective combination
of the RPDA and CE regimes \cite{DCE}. Also the dependence of ion
acceleration on different parameters of interaction were reported in
Refs. \cite{Petrov}.

In this paper we study the DCE regime of ion acceleration in the
intense laser pulse interaction with ultra-thin double layer (high
Z/low Z) \cite{hadron therapy,EsirkepovPRL,Schwoerer} mass limited
targets. The DCE regime comes into play when the laser pulse is able
not only to evacuate the electrons from the irradiated spot, but
also to accelerate the remaining ion core by the radiation pressure.
In this case the ion core experiences Coulomb Explosion while being
set in motion. This means that the ion core transforms into an ion
cloud expanding predominantly in the laser pulse propagation
direction. There is a charge separation field moving in front of the
positively charged ion cloud. In this field the protons of the
second layer are accelerated. In the present paper we study the
effect of mass limited targets on the proton acceleration because in
this case the influence of the electron return current, that can
compensate the positive charge in the irradiated spot, is reduced.
Also the evacuation of the electrons and the acceleration of the ion
core are more effective. We also consider the effects of different
laser pulse polarizations. We compare the acceleration by the
linearly polarized pulses with the acceleration by circularly
polarized ones. We optimize the acceleration process in both cases
varying the thickness and radius of the target.

The paper is organized as follows. In section 2 we review the
properties of the DCE regime of ion acceleration. The results of 2D
PIC simulations are presented in section 3. We conclude in section
4.

\section{The Directed Coulomb Explosion regime of ion acceleration.}

{\noindent} In this section we review the DCE regime of ion
acceleration from ultra-thin double-layer (high Z/low Z) foils of
solid density described in Ref. \cite{DCE}. The principal scheme of
the laser-foil interaction is shown in Fig. 1. In contrast to the
TNSA or CE or any other mentioned above scheme of ion acceleration
the DCE regime comes into play as an effective combination of
radiation pressure and coulomb explosion. It happens when the laser
pulse is able not only to expel the electron from the focal spot but
also to accelerate the remaining ion core by the radiation pressure.
Then the moving ions experience Coulomb explosion and the protons
from the second layer are accelerated by the moving charge
separation electric field. The proton energy can be estimated in
three steps as was shown in Ref. \cite{DCE}. First, the velocity of
the foil that is due to the radiation pressure should be calculated.
Then the energy gain due to the Coulomb Explosion in the moving with
the foil velocity reference frame should be obtained. The final
estimate is made by transforming back to the laboratory frame.

\begin{figure}[ht]
\epsfxsize5cm\epsffile{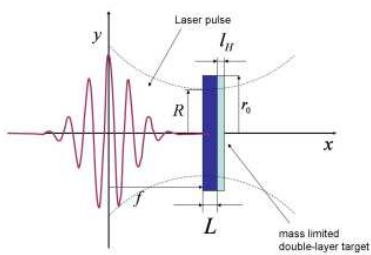} \caption{(color on-line) The
principal scheme of laser pulse interaction with a mass limited
double-layer solid density ultra-thin target.}
\end{figure}

Let us estimate the proton energy gain in DCE regime of proton
acceleration from double-layer target in the nonrelativistic case.
The radiation pressure is proportional to the intensity of the
pulse, $I\sim E_L^2$, where $E_L=E_{L}(t-x(t)/c)$ is the laser pulse
electric field. The total momentum gain due to the radiation
pressure is $P\sim E_{L}^2\tau\sim W/R^2$, here $\tau$ is the
duration of the laser pulse, $W$ is the laser pulse energy and $R$
is the radius of the focal spot. If we normalize the total momentum
of the accelerated fraction of the foil on the number of ions in
this fraction we obtain for the momentum of accelerated ions
$p_i/{m_ic}=W/N_i m_i c^2$, where $N_i=\pi R^2 L n_i$ is the number
of ions in the accelerated fraction. Here $n_i$ is the ion density
of the foil.

In the moving reference frame ($V/c=p_i/m_i c$) the protons will
acquire the energy $\mathcal{E}^\prime=mv^2/2$ due to the
longitudinal charge separation electric field produced by the heavy
ion core. $\mathcal{E}^\prime$, can be estimated as the energy gain
by a charged particle accelerated near the surface of a charged disk
and the acceleration distance is of the order of the disk radius,
$r_0$:
\begin{equation}
\mathcal{E}^\prime=2\pi^2
m_ec^2\frac{n_e}{n_{cr}}\frac{Lr_0}{\lambda^2}.
\end{equation}
Here $n_{cr}=\omega^2 m_e/4\pi e^2$ is the critical plasma density,
$\omega$ is the laser frequency.

The velocity in the laboratory frame is $V_{lab}=v+V$ and energy of
the accelerated protons reads as follows
\begin{equation} \label{proton energy}
\mathcal{E}=\mathcal{E}^\prime+\frac{p_i}{m_ic}\sqrt{2\mathcal{E}^\prime
m_pc^2}.
\end{equation}
As it was shown in Ref. \cite{DCE}, for a 500 TW laser pulse
interacting with a $0.1\lambda$ thick, $n_e=400 n_{cr}$ aluminum
foil with a layer of hydrogen on the back this formula gives a 100\%
energy increase over the value acquired in the static charge
separation field. A more detailed analysis of the acceleration
process was carried out in Ref. \cite{DCE} based on the results of
Ref. \cite{LP} for the RPDA regime. In Ref. \cite{DCE} the solution
of the equation of the foil motion under the action of the radiation
pressure is given, and the energy gain of protons due to DCE is
obtained in general case. In the nonrelativistic case the formula
(5) of Ref. \cite{DCE} reduces to Eq. (\ref{proton energy}) of the
present paper.

We should mention here that in order to expel all the electrons and
achieve the coulomb explosion the following condition on laser
electromagnetic vector potential
$a=0.85[I(\text{W/cm}^2)\lambda^2(\mu\text{m})10^{-18}]^{1/2}$ and
foil thickness L must be satisfied \cite{DCE}:
\begin{equation}
a=\pi\frac{n_e L}{n_{cr}\lambda}.
\end{equation}
From this condition one can see that for circularly and linearly
polarized pulses of the same total EM energy different foil
thicknesses follow. It is due to the fact that the value of vector
potential for linearly polarized pulse is square root of two times
bigger then that of the circularly polarized one. Though this will
lead to the reduction of the proton energy gain due to the coulomb
explosion (see Eq.(2)), it will also lead to the increase of foil
velocity due to radiation pressure (see Eq.(1)). The estimates
carried out based on Eq.(4) show a 10-20\% difference between the
maximum energy of protons accelerated in either circularly or
linearly polarized pulse interaction with foils of different
thicknesses. In order to obtain this estimate we also varied the
radius of the irradiated spot.

The utilization of the mass limited targets leads to different
effects on the effectiveness of proton acceleration in the DCE
regime. First of all, such targets provide a way to have a clean DCE
regime without the influence of the return current. Also the smaller
is the radius of the target the more monoenergetic the protons
become. However the reduction of the target size leads to the
reduction of the charge separation field and thus to the reduction
of $\mathcal{E}^\prime$. For the mass limited target with the radius
smaller then the focal spot radius, $R$, we will have the following
estimate
\begin{equation}
\mathcal{E}_{r_0>R}=\mathcal{E}^\prime(R/r_0)
+(p_i/m_ic)\sqrt{2\mathcal{E}^\prime m_pc^2}\sqrt{R/r_0}.
\end{equation}
We see that the maximum proton energy increases with the increase of
the target radius and reaches the value obtained in Eq. (4) for
$r_0=R$. However when the radius of the of the target exceeds the
focal spot radius the effects of charge compensation by the return
current come into play effectively reducing the estimate of the
maximum proton energy made in Eq. (4). That is why the mass limited
target with the radius of about the focal spot radius should
maximize the accelerated proton energy. In the next section we will
test this estimate against the results of 2D PIC simulations.

\section{2D PIC simulation results}

Here we report on the results of the 2D PIC simulations of an
intense laser pulse interaction with a double layer mass limited
target. We use the 2D PIC code REMP (relativistic electromagnetic
particle), which is a mesh code based on particle-in-cell method
\cite{Esirkepov_code}. The grid mesh size is $\lambda$/200, space
and time scales are given in units of $\lambda$ and $2\pi/\omega$,
respectively, the simulation box size is $20\lambda\times
10\lambda$. The 500 TW (peak intensity of $3 \times 10^{22}$
W/cm$^2$) 30 fs duration laser pulse, introduced at the left
boundary and propagating along the x axis from left to right, is
tightly focused ($f/D=1.5$, spot size is $1\lambda$ full width at
half maximum, FWHM) at the foil, which is placed at the distance of
$f=3.33\lambda$ from the left boundary. The pulse is either linearly
(along the z axis) or circularly  polarized. The target is composed
of two layers: high Z, fully ionized silicon Si$^{+14}$ with an
electron density of $600 n_{cr}$, thickness $L$; and low Z layer,
ionized hydrogen (H$^+$, $n_H=30 n_{cr}$), with a thickness $l_H$.
As mentioned above, the material of the foil is assumed to be fully
ionized. This is justified since the intensity needed to fully
ionize silicon is approximately $10^{21}$ W/cm$^2$, while we use in
simulations the intensity at least one order of magnitude larger. So
the foil will become fully ionized well before the arrival of the
pulse peak and can be modeled as a double layer slab of the
overdense plasma.

\begin{figure}[ht]
\begin{tabular}{cc}
\epsfxsize8cm\epsffile{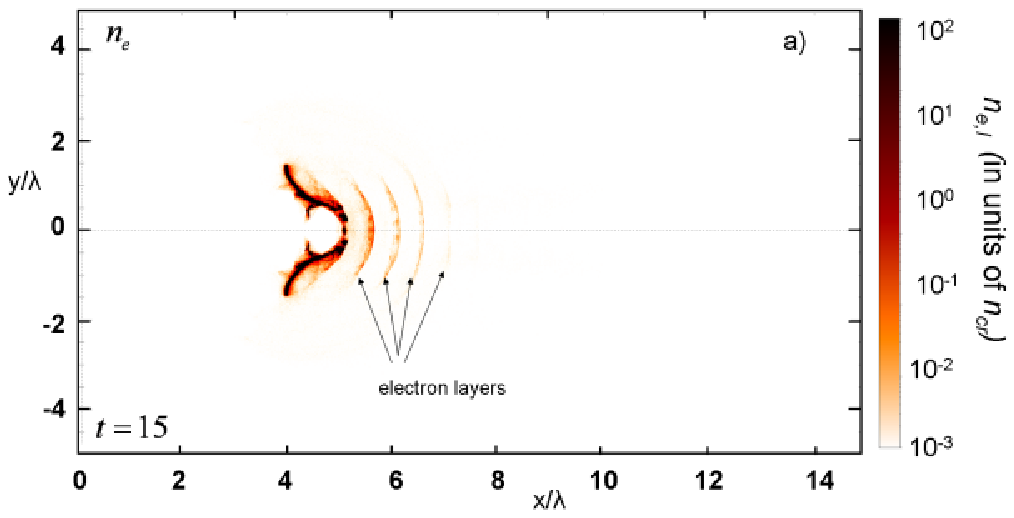} &
\epsfxsize8cm\epsffile{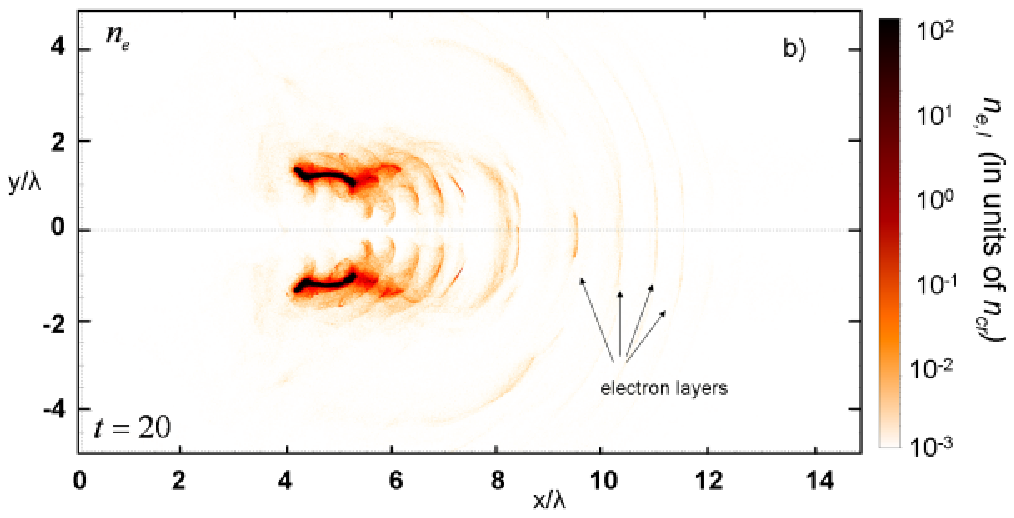}
\end{tabular} \caption{(color on-line) The distributions of electron
density at $t=15$ (a) and $t=20$ (b) for a 500 TW linearly polarized
laser pulse interaction with a double layer (Si+H) mass limited
target, $R=1.5\lambda$.}
\end{figure}

In order to illustrate the interaction of a laser pulse with mass
limited target we present Fig. 2 where the distribution of electron
density for two time instants is shown. One can see that under the
action of intense laser pulse the electrons that are not in the
focal spot are pushed aside, and those that are in the focal spot
are evacuated in the form of thin layers, as was pointed out in Ref.
\cite{KPopov}. These layers, flying with the velocity approaching
the speed of light, can be utilized as relativistic mirrors to
create ultra-short electromagnetic pulses from the
counterpropagating pulses as was shown in Ref. \cite{Bulanov_atto}.
This is an illustration that relativistic mirrors can be created not
only in the laser pulse interaction with gaseous targets but also in
the interaction with solid density targets, in the regime that is
considered in regard with the ion acceleration. The formation of
these layers can clearly be seen in the electron distribution in
phase space (see Fig. 3).

\begin{figure}[ht]
\begin{tabular}{cc}
\epsfysize5cm\epsffile{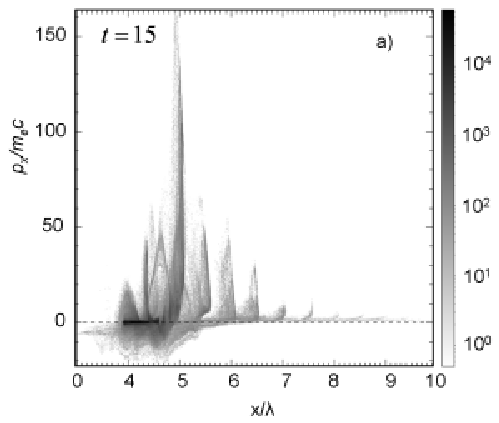}&\epsfysize5cm\epsffile{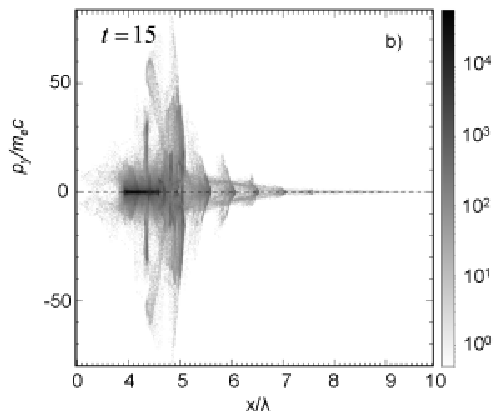}
\end{tabular}
\caption{The distribution of electrons in phase space $(x,p_x)$ (a)
and $(x,p_y)$ (b) at $t=15$ for a 500 TW linearly polarized laser
pulse interaction with a double layer (Si+H) mass limited target,
$R=1.5\lambda$.}
\end{figure}

At the time when the electrons are pushed from the focal spot, the
heavy ions of the first layer begin to move under the action of the
radiation pressure, see Fig. 4a. While in motion the heavy ion layer
experience Coulomb explosion due to the excess of positive charge.
It leads to the transformation of the heavy ion layer into the
expanding predominantly in the direction of laser pulse propagation
ion cloud, see Fig. 4b. This expanding cloud generates a charge
separation longitudinal electric field (Fig. 5), which accelerates
the protons of the second layer.

\begin{figure}[ht]
\begin{tabular}{cc}
\epsfxsize8cm\epsffile{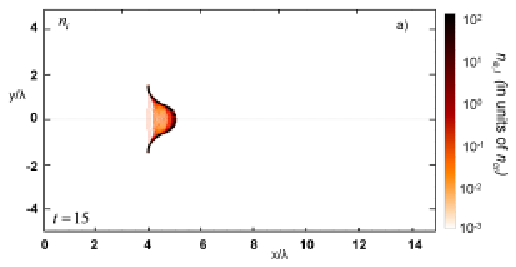} & \epsfxsize8cm\epsffile{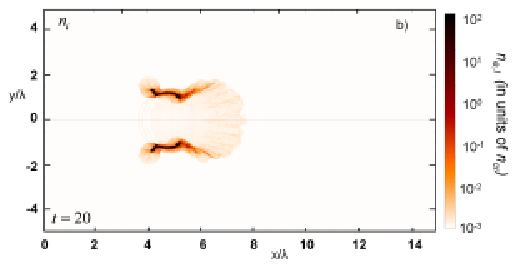}
\end{tabular} \caption{(color on-line) The distribution ion density
at $t=15$ (a) and $t=20$ (b) for a 500 TW linearly polarized laser
pulse interaction with a double layer (Si+H) mass limited target,
$R=1.5 \lambda$.}
\end{figure}

\begin{figure}[ht]
\epsfxsize10cm\epsffile{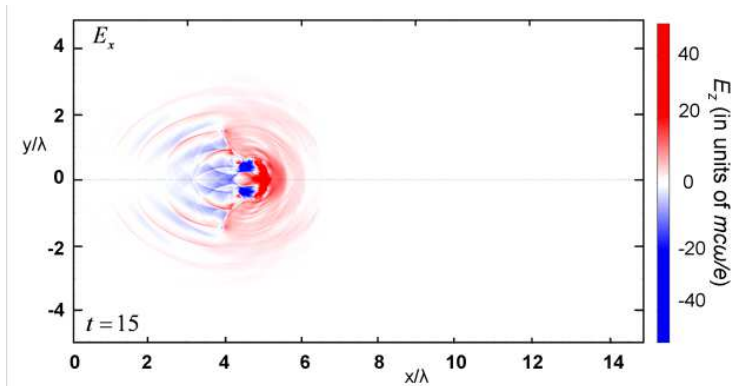} \caption{(color on-line) The
distribution of the x-component of electric field at $t=$ for a 500
TW linearly polarized laser pulse interaction with a double layer
(Si+H) mass limited target, $R=1.5\lambda$.}
\end{figure}

The protons first experience the acceleration due to the radiation
pressure and then the longitudinal charge separation field comes
into play. The protons are accelerated in a form of a thin layer
flying in front of the expanding heavy ion cloud almost in all
directions (Fig. 6). However the most energetic protons, as can be
seen from the proton distribution in $(p_x,p_y)$ plane (Fig. 7a),
are going in the laser propagation direction.

\begin{figure}[ht]
\begin{tabular}{cc}
\epsfxsize8cm\epsffile{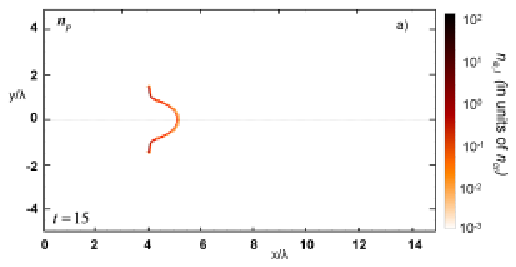} &
\epsfxsize8cm\epsffile{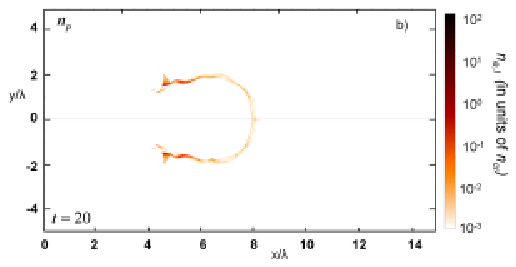}
\end{tabular} \caption{(color on-line) The distribution proton
density at $t=15$ (a) and $t=20$ (b) for a 500 TW linearly polarized
laser pulse interaction with a double layer (Si+H) mass limited
target, $R=1.5\lambda$.}
\end{figure}

In order to compare the cases of linear and circular polarizations
we show the distribution of protons in $(p_x,p_y)$ plane in both
cases. It can be easily seen that the maximum energy of protons is
almost the same, as well as the angular distribution. The larger
thickness of the proton distribution in the case of circular
polarization should lead to a broader peak in proton spectrum.

\begin{figure}[ht]
\begin{tabular}{cc}
\epsfysize5cm\epsffile{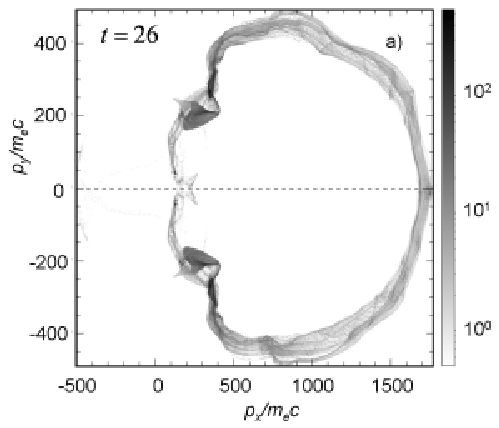}&\epsfysize5cm\epsffile{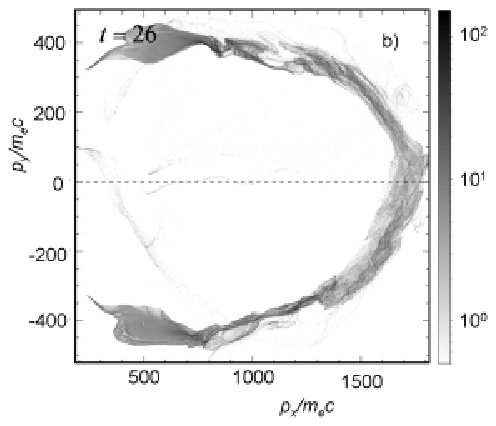}
\end{tabular}
\caption{The distribution of protons in $(p_x,p_y)$ plane
accelerated by linearly (a) and circularly (b) polarized laser
pulses.}
\end{figure}

The spectra of accelerated protons in the cases of circular and
linear polarization of the laser pulse are shown in Fig. 8. These
protons are contained inside an angle of $5^o$ from the target
normal. Here a 500 TW tightly focused (f/D=1) laser pulse interacted
with a $0.1\lambda$ thick disk with the radius of $1.5\lambda$ for
linear and $1.0\lambda$ for circular polarization. Both spectra have
maxima around $\mathcal{E}=320$ MeV and extend to the maximum energy
of $\mathcal{E}=370$ MeV. However the protons accelerated by the
linearly polarized pulse demonstrate a monoenergetic behavior with a
sharp peak in spectrum with $\Delta\mathcal{E}/\mathcal{E}\sim 1\%$.
Whereas in the case of circular polarization of the laser pulse the
protons have a wide distribution with
$\Delta\mathcal{E}/\mathcal{E}\sim 25\%$. This could be expected
from the analysis of the proton distribution in $(p_x,p_y)$ plane.

\begin{figure}[ht]
\epsfxsize10cm\epsffile{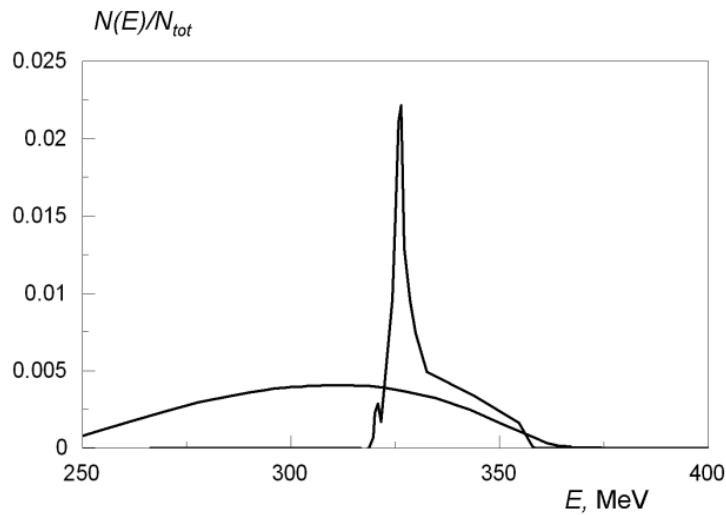} \caption{The spectra of protons
accelerated inside an angle of $5^o$ to the target normal at $t=26$
for 500 TW linearly and circularly polarized laser pulses
interacting with a double layer (Si+H) mass limited target,
$R=1.5\lambda$ for linear polarization and $R=1 \lambda$ for
circular polarization.}
\end{figure}

As it was mentioned above the proton maximum energy depends on the
size of a mass limited target. In Fig. 9 we present several
dependencies of maximum proton energy on the radius of the target
for different target thicknesses and different laser polarizations.

\begin{figure}[ht]

\epsfysize8cm\epsffile{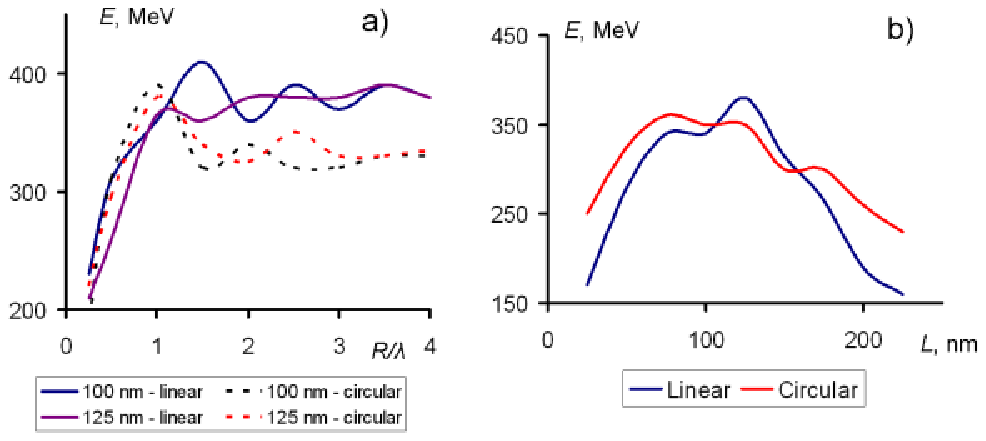} \caption{The dependencies of
the proton maximum energy (a) on target radius for different target
thicknesses and (b) on target thickness for different laser
polarizations for a 500 TW laser pulse interaction with a double
layer (Al+H) mass limited target}
\end{figure}

However all these curves has one feature in common. At first, the
maximum proton energy steadily increases with the increase of the
transverse size of the target. They have a maximum around
$R=\lambda$, \textit{i.e.} when the target is about the size of the
focal spot. Further increase of the target size affects the maximum
proton energy only slightly. This result can be expected from the
analytical estimate, carried out in the previous section.

We should also note that the absolute maxima of $\mathcal{E}_{max}$
for circular and linear polarizations are very close. However they
clearly correspond to different target thicknesses. As it was
mentioned in the previous section is due to the fact that for
circular polarization the optimal target thickness is smaller then
for linear one since $a$ in the case of circular polarization is
smaller.

\section{Conclusion.}

In this paper we studied the Directed Coulomb Explosion regime of
ion acceleration from double-layer solid density mass limited
targets. As it was shown in Refs. \cite{DCE} this regime which is an
effective combination of the radiation pressure and coulomb
explosion mechanisms of ion acceleration allows for the production
of quasimonoenergetic protons beams which are highly sought for a
number of applications. In this regime the laser pulse not only
removes the electrons from the irradiated area but also accelerates
the remaining ion core. Thus this core is transformed into an
expanding predominantly in the direction of the laser pulse
propagation ion cloud. This positively charged cloud is the source
of the longitudinal electric field that accelerated the light ions
from the second layer.

Here we focused on the study of DCE regime of acceleration in laser
pulse interaction with mass limited targets. Such targets allow for
the reduction of the electron return current that can compensate the
positive charge in the irradiated spot. Also the use of mass limited
target increases the effectiveness of electron expulsion from the
focal spot and the monoenergeticity of accelerated protons. We
studied the dependency of the proton maximum energy on the
transverse size of the target. As it increases so does the maximum
energy of protons due to the fact that the energy gain in Coulomb
field increases while the energy gain due to radiation pressure
stays the same. When the radius of the target reaches the same size
as the focal spot radius this increase stops. Further increase of
the target transverse size leads to the slow reduction of the
maximum proton energy in most cases which can be attributed to the
effect of the return current. That is why we argue that the mass
limited targets with the transverse size matching the focal spot of
the laser pulse are most favorable for the proton acceleration in
the DCE regime. However this statement is true for the targets
having optimal thicknesses.

We were also interested in the effect of laser pulse polarization on
the maximum proton energy due to the fact that recently a number of
papers appeared claiming the advantages of using the laser pulses
with circular polarization for ion acceleration. We indeed find that
the optimum conditions for ion acceleration differ for circular and
linear polarizations. It is however is connected with the fact that
for circular polarization the maximum achievable value of the
electric field is smaller then that for linear polarization. That is
why the optimal target thickness should be smaller in the former
case. We should note here that smaller target thickness does not
necessarily leads to the reduction of the maximum proton energy
because of the reduction of the coulomb repulsion field of the bare
ion core. As the energy gain due to the Coulomb Explosion decreases,
the energy gain due to the Radiation Pressure increases. That is why
we did not find a strong difference between the maximum proton
energies obtained in either circularly or linearly polarized pulse
interactions with solid density foils in the DCE regime of ion
acceleration for a wide range of parameters.

\acknowledgments

This work was supported by the National Science Foundation through
the Frontiers in Optical and Coherent Ultrafast Science Center at
the University of Michigan. The authors would like to thank Dr. T.
Zh. Esirkepov for providing REMP code for simulations. The authors
acknowledge fruitful discussions with Prof. V. Yu. Bychenkov.

\end{document}